\documentclass[a4paper]{article}

\usepackage[english]{babel}
\usepackage[utf8]{inputenc}
\usepackage{amsthm}
\usepackage{amsmath}
\usepackage{amsfonts}
\usepackage{amssymb}
\usepackage{hyperref}
\hypersetup{
hidelinks,
colorlinks=false,
linkcolor=black,
citecolor=black
}
\usepackage{caption}
\usepackage{subcaption}
\usepackage{enumitem}
\usepackage{booktabs}
\usepackage{float}
\usepackage{bm}
\usepackage[numbers]{natbib}
\usepackage{appendix}
\usepackage{graphicx}
\usepackage{utfsym}
 
\usepackage{comment}
\usepackage{setspace}
\renewcommand{\baselinestretch}{1.2}
\def\v{\mathcal{V}}
\def\e{{\mathrm{E}}}
\def\ii {{\mathcal{I}}}
\def\g{{\mathcal{G}}}
\def\bfr{{\bm{R}}}
\def\hatrho{{\hat{\rho}}}
\def\bfx{{\bm{X}}}

\def \det {\operatorname{det}}
\def\tildew{\widetilde{w}}
\def\lij {L_{\{i,j\}}}
\def\tildel{\Tilde{L}}

\def\diag{\text{diag}}

\def\sigb{{\boldsymbol{\Sigma}}}

\def\Cov{{\rm Cov}\,}
\def\rt#1{\sqrt{#1}\,}

\usepackage{tikz}
\usepackage{authblk}

\makeatother

\newcommand{\blind}{0}

\begin{document}
\def\spacingset#1{\renewcommand{\baselinestretch}%
{#1}\small\normalsize} \spacingset{1}

\spacingset{1.45} 

\date{February, 2023}

\if0\blind
{
  \title{\bf Vine dependence graphs with latent variables as summaries for gene expression data}
  \author[1]{Xinyao Fan}
\author[1]{Harry Joe}
\author[1]{Yongjin Park}

\affil[1]{Department of Statistics, University of British Columbia, Canada}

  \maketitle
} \fi

\begin{abstract}
The advent of high-throughput sequencing technologies has
lead to vast comparative genome sequences. The construction of gene-gene
interaction networks or dependence graphs on the genome scale is vital
for understanding the regulation of biological processes.  Different
dependence graphs can provide different information.

Some existing methods for dependence graphs based on high-order
partial correlations are sparse and not informative when there are
latent variables that can explain much of the dependence in groups of
genes. Other methods of dependence graphs based on correlations and
first-order partial correlations might have dense graphs.  When genes
can be divided into groups with stronger within group dependence in
gene expression than between group dependence, we present a dependence
graph based on truncated vines with latent variables that makes use of
group information and low-order partial correlations. The graphs are not dense, and the genes that might be more central have
more neighbors in the vine dependency graph.  We demonstrate the use of
our dependence graph construction on two RNA-seq data sets --- yeast
and prostate cancer. There is some biological evidence to support the
relationship between genes in the resulting dependence graphs.

A flexible framework is provided for building dependence graphs 
via low-order partial correlations and formation of groups, leading
to graphs that are not too sparse or dense.  We anticipate that this
approach will help to identify groups that might be central to different
biological functions.

\textbf{Keywords}: {partial correlation}, {Gaussian factor model}, {dependence graph}, {gene-gene network}, {truncated vine}, {latent dependence}, {RNA-seq}
\end{abstract}

\section{Introduction}

With the rapid advancements in accuracy and throughput, high-throughput sequencing technologies enabled researchers to routinely measure mRNA transcript levels over tens of thousands of genes. Since observed gene expression values were derived from interactions between genes, having gene expression profiles across many samples has naturally sparked scientific interest in inferring such interaction patterns from the observed data. Gene-gene interaction networks implicate functional organizations of complex biological of biological pathways and essential cellular processes. Building an accurate interaction network of genes improves our understanding of complex biological systems and provides a tool to make proper interpretations of large, high-dimensional genomics data. Since the invention of microarray technology \cite{schena1995quantitative}, RNA-sequencing \cite{wang2009rna}, and single-cell RNA-sequencing \cite{macosko2015highly} to date, finding an underlying network structure from observed data has been a fundamentally important problem in biology.




Many approaches have been proposed in statistics to understand interaction patterns embedded in a large-scale, high-dimensional data matrix by constructing a statistical (conditional) dependency graph. Traditionally, the inference problem has been formulated to search for an optimal model in the class of Gaussian graphical models or other relevant conditional independence graphs \cite{Whittaker1990}. There, we measure \emph{high-order} partial correlation values directly from the inverse covariance matrix estimated from observed data and identify one type of dependence graph. However, finding such a complex dependency graph could become intractable in computation, resulting in a graph model overfitting to observed data stringent regularization penalties were imposed, see \cite{lin2016estimation} and references therein. Alternatively, dependence graphs can be constituted of \emph{low-order} partial correlations, which would only rely on relatively fast and reliable estimates. The examples include a traditional spanning tree model \citep{chow1968approximating} and a vine structure \citep{kurowicka2003parameterization}. Although such a method based on low-order partial correlation patterns generally leads to a more informative dependence graph than the high-order graphs given the limited sample size \citep{magwene2004estimating}, we noticed challenges in real-world data, which were generally affected by a handful of latent group-wise effects. Unless we properly address latent group variables, which explain a large proportion of variance in the observed data, we show that learning a conditional independence graph is not suitable and may lead to an unwanted sparse graph.


This work will propose a new methodology for inferring a dependency graph from gene expression data, building on a \emph{low-order} vine graph model \citep{kurowicka2003parameterization} while incorporating latent group variables to improve the resolutions among the observed variables along with the relationships with the nuisance variables introduced by a latent group structure.



\section{Summary of existing methods}
\label{sec-methods}
In this section, a few methods for constructing dependence graphs are summarized. The methods
are defined in terms of partial correlations computed from a covariance or correlation matrix so the main formulas for partial correlations are given in section \ref{sec-partialcor} before methods are summarized
in Section \ref{dependence-graphs}. The methods include the conditional independence graph with thresholding, the FOCI method in \cite{magwene2004estimating} and the truncated partial correlation vine. A more basic graph is based on the Thresholding-Correlation method, for which two variables are connected in the dependence graph if their absolute correlation exceeds a given threshold. Section  \ref{sec-comparisons} includes the simple case of a correlation
matrix based on a factor structure to illustrate that the
Thresholding-Correlation method tends to be too dense and the conditional
independence graph tends to be too sparse when the dimension $d$ is large.

\subsection{Partial correlations}
\label{sec-partialcor}
This section summarized some required results on partial correlations in order to explain different
methods of constructing dependence graphs.
There is an assumption that variables jointly have a multivariate Gaussian distribution after appropriate transforms on each variable.

Let the (transformed) observation vector be $\bfx_{\ii}=(X_1,\ldots,X_{d})$, where $d\geq 2$ is large and $\ii$ is the index set $\{1,2,\ldots,d\}$; Suppose $\bfx_{\ii}$ has multivariate Gaussian distribution, denoted $\bfx_{\ii}\sim N(\mu,\sigb)$, where $\sigb$ is the $d\times d$ covariance matrix. Let $S \subseteq \ii $ and have at least cardinality 2, i.e. $|S|\geq 2$. For a pair $i,j \in S$, $i\neq j$, denote $S$ with $\{i,j\}$ removed by $\lij:=S_{-\{i,j\}}=S\backslash \{i,j\}$. Let the corresponding random vector be $\bfx_{\lij}:=\{X_{k},k\in \lij \}$. Define $\rho_{ij;\lij}$ as the partial correlation as the correlation parameter of the conditional (Gaussian) distribution of $[X_i,X_j|\bfx_{\lij}]$. It quantifies the dependence between $X_i$ and $X_{j}$ without the linear effect of $\bfx_{\lij}$.

The cardinality  of  $\lij$ refer to the order of the partial correlation coefficient. In the case where order is zero, $|S|=|\{i,j\}|=2$ and $\lij=\emptyset$. Then $\rho_{i,j;\emptyset}=\rho_{ij}$, which is the standard pairwise correlations between $X_{i}$ and $X_{j}$, $i,j\in \ii, i\neq j$. For any order $|\lij|$ with $|\lij|>1$, \cite{anderson1958introduction} derives a formula to recursively calculate the partial correlations in terms of lower-order (partial) correlations.
Suppose $|S|\subseteq \ii$, with cardinality $|S|\geq 3$. Consider a set $\{i,j,k\}\subseteq S$ with three distinct indices. Let $\tildel:=S_{-\{i,j,k\}}$ and $\lij=\tildel \cup k$. Then 
\begin{equation}
\label{partial-correlations}
\rho_{i,j;\lij}=\frac{\rho_{i,j;\tildel}-\rho_{i,k;
\tildel }\rho_{j,k;\tildel}}{\sqrt{1-\rho_{i,k;\tildel}^2}
\sqrt{1-\rho_{j,k;\tildel}^2}}
\end{equation}

Let the precision matrix (inverse of covariance matrix) $\sigb^{-1}=(\sigma^{ij})$ and $\ii_{ij}=\ii \backslash\{i,j\}$. 
For the inverse correlation matrix, it is well known from \cite{Whittaker1990} section 5.8 that for $i\neq j$,
\begin{equation}
\label{eq-partcor-given}
    \rho_{ij;\ii_{\{ij\}}}=-\frac{\sigma^{ij}}{\sqrt{\sigma^{ii}\sigma^{jj}}}
\end{equation}
These are the highest order partial correlations.

\subsection{Summary of dependence graphs}
\label{dependence-graphs}

\noindent\textbf{Conditional dependence graph with thresholding}\\
Let $\v=\{1,\ldots,d\}$ be the set of vertices, one for each variable.
The edge set $\e$ consist of unordered pairs $(i,j)$, $(i,j)\in \e$ if
there is an edge between $X_i$ and $X_j$.  For a theoretical
covariance matrix, $X_i$ and $X_{j}$ are connected in the graph if
$X_{i}\not\perp X_{j}\big\vert X_{\lij}$ and are not connected in the
graph only if $X_{i}\perp X_{j}\big\vert X_{\lij}$. That is, there is
no edge connected variables $i,j$ if $\sigma^{ij}=0$ or equivalently
$\rho_{ij;\lij}=0$ for $i\neq j$.

For a sample covariance matrix, Section 6.1 of \cite{Whittaker1990}
suggests a naive procedure with a threshold.  The sample-based
conditional independency graph is based on the rule that an edge for
two variables is \textit{not} included in the graph if the absolute
high-order partial correlation coefficient in
\eqref{eq-partcor-given} for these two variables is below the
threshold.

\vspace{0.2cm}

\noindent\textbf{FOCI (first-order-conditional independence method) }\\
In the FOCI method proposed in \cite{magwene2004estimating}, the two variables $X_i$ and $X_j$ are connected only if there are no other variables in the analysis for which $X_i$ and $X_j$ are conditionally independent or which causes an association reversal (that means partial correlation given some variable $X_k$ and
correlation between $X_i$, $X_j$ has opposite signs). Their method defines a modified first-order partial correlation similar to equation \eqref{partial-correlations}. 

More formally, $e_{ij}$ is an edge in $\g$, if and only if there is no other variable, $k$ ($k\neq i\neq j$) such that $\hatrho_{ij;k}\approx 0$ or $\hatrho_{ij;k}<0$ where
\begin{equation*}
    \hatrho_{ij;k}=\frac{|\rho_{ij}|-|\rho_{ik}||\rho_{jk}|}{\sqrt{1-\rho_{ik}^2}\sqrt{1-\rho_{jk}^2}}
\end{equation*}
A threshold is used for the closeness to 0.

\vspace{0.2cm}

\noindent\textbf{Truncated partial correlation vines}\\
\cite{bedford2002vines} propose vines as a graphical object to summarize conditional dependence; definitions for vine are in a later Section \ref{sec-vine-concept}.
A common use is to find a parametrization
of the correlation matrix so that dependence can be summarized with low-order partial correlations;
that is, higher-order partial correlations are small in absolute value. An algorithm to do
this is briefly summarized next.

First, a maximal spanning tree with $(d-1)$ edges is constructed combining variables with the
largest absolute correlations. The tree can summarize all of the dependence of a set of variables only if A-B-C is a segment of the tree, then variables A and C are conditional independent given variable B. That is, if a maximal spanning tree summarizes the dependence, then there is Markov dependence meaning that two variables are conditionally independent given the variables in the path between the two. If the correlation matrix based on this tree is not close to the (empirical) correlation matrix, then additional edges of 2 variables are added based on large absolute partial correlation given one conditioning variable (2 variables are each connected to the conditioning variable).
Subsequently more edges can be added based on conditioning on 2 or more variables. More details about vines are reviewed in Section \ref{sec-vine-concept}.

\subsection{Comparisons}
\label{sec-comparisons}

Some comments of the different dependence graphs are summarized in this section. A numerical
example is given to show that the conditional independence graph with thresholding can lead to
sparse graphs when latent variables can explain much of the dependence among the variables.
A theoretical explanation is given for this property.

\begin{itemize}
    \item Thresholding on correlations: 
    The main problem with this method is that it leads to a dense graph if the variables are mostly at least moderately correlated with each other. For high-dimensional data, if there can be a cluster or subset of variables that are highly correlated, the dependence graph will have a near-complete graph for this cluster. In addition, the method does not consider any information on conditional dependence.
    \item Conditional dependence graph with thresholding:
    
   The method relies on nonzero elements in the inverse correlation matrix
  related to the partial correlation of two variables given the
  remaining variables.  However, the high-order interactions in the
  graph are usually not easy to interpret.  For high-dimensional
  correlation matrices based on structured factor models, there are
  theoretically no zeros in the inverse correlation matrix if each
  variables loads to at least one latent factor; see for example the
  form of the inverse correlation matrix in Section 3.10.4 of
  \cite{Joe2014}.  But as the dimension gets larger and larger, many
  entries of the inverse correlation matrix approach zero.  Consider a
  1-factor model and assume the dependence in the model is
  strong. \citep{fan2022high} show that in such case, the latent
  variable can be recovered consistently from the observed variables
  as the dimension goes to infinity.  This indicates that the
  conditional correlation of two variables given the remaining
  variables is asymptotically equivalent to the conditional
  correlations of two variables given the latent variables; the latter
  conditional correlation is zero.  Given a threshold, the dependency
  graph tends to be sparser as the number of variables increases if
  added variables continue to load moderately on the latent variable.
  Therefore, constructing graphs based on non-zeros elements in the
  precision matrix is not informative for the Gaussian factor
  dependence structures when the dimension is large.
    
  The conditional dependency graph can be useful when $\sigb$ comes
  from models not linked to latent variables.  One example is from
  Proposition 2 of \cite{joe2018parsimonious} where $\sigb$ is based
  on a truncated partial correlation vine.  The conditional
  independency graph based on $\sigb^{-1}$ are parsimonious and
  informative.  If the vine truncation level is $m$, the edges of the
  resulting conditional independency graph are those in trees 1 to $m$
  of the vine; these involves $\sum_{i=1}^m (d-i)$ non-zero entries in
  $\sigb^{-1}$.  The remaining $(d-m)(d-m-1)/2$ positions of the
  $\sigb^{-1}$ are zero.

    \item FOCI: This method can be considered as improvement on the direct threshold method by
    considered first-order partial correlations. However, the method only consider the (modified) first-order partial correlation. For variables that are highly correlated due to several latent variables (assume only positive dependence exists), it will produce a dense connected graph, because information from higher-order partial correlation is not used.

\item Truncated partial correlation vines: 
The partial correlation vine is explained in details in Section \ref{sec-vine-concept}. Before details of the dependence graph based on the vine is given, the example
below is used to show that, for factor structures, it avoids the denseness of the graph based
on thresholding correlations and it avoids the sparseness of the conditional independence
graph. The reason is that it makes good use of the strong correlations and some additional
low-order partial correlations. 

In the dependence graph based on a vine, the first tree is a maximal spanning tree and its
edges are shown in the color black with the correlation labeling the edges between pairs of
variables. Additional coloured edges are drawn to indicate some partial correlations that exceed thresholds. Blue, red, and green colored edges are used to represent the partial
correlations between variables given one, two, and more variables respectively.

\end{itemize}

\textbf{Examples to compare different methods}
A small illustrative example is provided to compare different dependence graphs built by various methods. Data are simulated from the
1-factor model with latent variable $W$:
\begin{equation}
    Z_{j}=\alpha_{j}W+\psi_{j}\epsilon_{j}\quad 
  j\in\{1,\ldots,D\},
  \label{eq-1factor-gauss}
\end{equation}  
where $W,\epsilon_1,\epsilon_2,\ldots$ are mutually independent $N(0,1)$
random variables, and $-1<\alpha_j<1$, $\psi_j^2=1-\alpha_j^2$
for all $j$. Let the loading matrix/vector be $A=[\alpha_{1},\ldots,\alpha_{d}]^{T}$, $\Psi^{2}=\diag(\psi_1^2,\ldots,\psi_{d}^{2})$. The correlation matrix of $(Z_1,\ldots,Z_{d})^{T}$ is $\sigb=AA^{T}+\Psi^{2}$ and the correlation matrix of $(Z_1,\ldots,Z_{d},W)^{T}$ is $\sigb^{*}=\begin{bmatrix}
\sigb &  A\\
A^{T} & 1\\
\end{bmatrix}.$ 

Suppose $d=10$ loadings are generated uniformly from interval $[0.5,0.9]$.
With weaker loadings, a  larger $d$ would be needed to illustrate
the comparisons.
In one simulation, after sorting into decreasing order,
the values in $A$ is $[0.90, 0.86, 0.81, 0.77, 0.72, 0.68, 0.63, 0.59, 0.54, 0.50]^{T}$. 
The four methods are applied to both $\sigb$ and $\sigb^{*}$; the former
matrix is a submatrix of the latter without the last rows and column. The correlation matrix $\sigb^{*}$ is:

$$
\begin{bmatrix}
1.00 & 0.77 & 0.73 & 0.69 & 0.65  & 0.61 & 0.57 & 0.53 & 0.49 & 0.45 & 0.90\\
0.77 & 1.00 & 0.69  & 0.66  & 0.62 & 0.58 & 0.54 & 0.50 & 0.47 &  0.43 & 0.86\\
0.73 & 0.69 & 1.00 & 0.62 & 0.59  & 0.55 & 0.51 & 0.48 &  0.44 &  0.41 & 0.81\\
0.69 & 0.66 & 0.62&  1.00 & 0.55  & 0.52   & 0.49 &  0.45  & 0.42  & 0.38 & 0.77\\
0.65 & 0.62 & 0.59 & 0.55  & 1.00&  0.49&  0.46 & 0.43 & 0.39 & 0.36 & 0.72\\
0.61  & 0.58 &  0.55 &  0.52 & 0.49 & 1.00 & 0.43 & 0.40 & 0.37 &  0.34 & 0.68\\
0.57 & 0.54  & 0.51  & 0.49 & 0.46 & 0.43 & 1.00  & 0.37  &0.34  & 0.32 & 0.63\\
0.53 & 0.50 & 0.48 & 0.45 & 0.43 & 0.40 & 0.37& 1.00 & 0.32 & 0.29 & 0.59\\
0.49 & 0.47 & 0.44 & 0.42 & 0.39 & 0.37&  0.34  & 0.32 &  1.00  & 0.27 & 0.54\\
0.45 &  0.43  & 0.41 & 0.38  & 0.36 & 0.34&  0.32 & 0.29  &0.27 &  1.00 & 0.50\\
0.9 & 0.86  & 0.81   & 0.77  & 0.72  & 0.68 
& 0.63 & 0.59 & 0.54  &  0.50 & 1.00\\
\end{bmatrix}$$
Using \eqref{eq-partcor-given}, the partial correlation matrix based on $\sigb$ is:
$$\begin{bmatrix}
1.00 & 0.29  & 0.24  & 0.20 & 0.17  & 0.15 & 0.13 & 0.12  & 0.11   & 0.09\\
0.29 & 1.00  & 0.18  & 0.15  & 0.13  & 0.11 & 0.10   & 0.09  & 0.08  & 0.07 \\
0.24 & 0.18 & 1.00  & 0.12 & 0.11 & 0.09 & 0.08 & 0.07 & 0.06  & 0.06\\
0.20 & 0.15 & 0.12  &1.00  &0.09 & 0.08 & 0.07 & 0.06 & 0.05 & 0.05\\
0.17 & 0.13  & 0.11 &. 0.09 & 1.00 & 0.07 & 0.06 & 0.05 & 0.05 &  0.04\\
0.15 & 0.11 &0.09 &0.08& 0.07& 1.00& 0.05& 0.05& 0.04  &0.04\\
 0.13 & 0.10  & 0.08 & 0.07  &0.06 & 0.05&  1.00 & 0.04  &0.04   &0.03\\
 0.12 & 0.09  & 0.07  & 0.06  & 0.05  & 0.05  & 0.04&  1.00 & 0.03 &  0.03\\
 0.11 & 0.08 & 0.06  &0.05 & 0.05 & 0.04 & 0.04 & 0.03  &1.00 &  0.03\\
 0.09  & 0.07 & 0.06  &0.05 &0.04 & 0.04&  0.03&  0.03 & 0.03 &  1.00\\
\end{bmatrix}
$$
and the partial correlation matrix based on $\sigb^{*}$ is:
$$\begin{bmatrix}
1.00  & 0.00 & 0.00 & 0.00  & 0.00  &0.00 & 0.00  & 0.00 & 0.00  & 0.00 &   0.53\\
0.00 & 1.00 &  0.00 & 0.00 & 0.00&  0.00 &0.00 &0.00 &0.00 & 0.00&  0.42\\
0.00 & 0.00 &  1.00 & 0.00 & 0.00&  0.00 &0.00 &0.00 &0.00 & 0.00&  0.36\\
0.00 & 0.00 &  0.00 & 1.00 & 0.00&  0.00 &0.00 &0.00 &0.00 & 0.00&  0.31\\
0.00 & 0.00 &  0.00 & 0.00 & 1.00&  0.00 &0.00 &0.00 &0.00 & 0.00&  0.27\\
0.00 & 0.00 &  0.00 & 0.00 & 0.00&  1.00 &0.00 &0.00 &0.00 & 0.00&  0.24\\
0.00 & 0.00 &  0.00 & 0.00 & 0.00&  0.00 &1.00 &0.00 &0.00 & 0.00&  0.21\\
0.00 & 0.00 &  0.00 & 0.00 & 0.00&  0.00 &0.00 & 1.00 &0.00 & 0.00&  0.19\\
0.00 & 0.00 &  0.00 & 0.00 & 0.00&  0.00 &0.00 & 0.00 & 1.00 & 0.00&  0.17\\
0.00 & 0.00 &  0.00 & 0.00 & 0.00&  0.00 &0.00 & 0.00 & 0.00 & 1.00&  0.15\\
 0.53 & 0.42 &0.36 &0.31& 0.27& 0.24 &0.21& 0.19& 0.17 & 0.15  &1.00\\
\end{bmatrix}$$

The connected edges of the 4 methods of dependence graphs are summarized in Table \ref{tab-example-edges}.
\begin{table}[H]
    \centering
    
    \begin{tabular}{c|c|c}
    \toprule
   method & connected edges & \# edges\\
   \hline
   $\sigb$ & &\\
   \hline
    TC   &  complete graph & 55\\
    \hline
    CDG   & 1--2, 1--3, 1--4 & 3\\
    \hline
    FOCI  & 1--2, 1--3, 1--4, 1--5, 1--6, 1--7, 1-- 8, 1--9, 1--10, 2--3, 2--4, 2--5, 2--6, 3--4, 3--5 & 15\\
    \hline
    Vine  & 1--2, 1--3, 1--4, 1--5, 1--6, 1--7, 1-- 8, 1--9, 1--10, 2--3, 2--4, 2--5, 2--6 &  13\\
   \midrule
    $\sigb^{*}$ & &\\
    \hline
    TC   &  complete graph & 66\\
    \hline
    CDG   & 1--$W$, 2--$W$, 3--$W$, 4--$W$, 5--$W$, 6--$W$, 7--$W$ & 7\\
    \hline
    FOCI  & 1--$W$, 2--$W$, 3--$W$, 4--$W$, 5--$W$, 6--$W$, 7--$W$, 8--$W$, 9--$W$, 10--$W$ & 10\\
    \hline
    Vine  &  1--$W$, 2--$W$, 3--$W$, 4--$W$, 5--$W$, 6--$W$, 7--$W$, 8--$W$, 9--$W$, 10-$W$ & 10\\
         \bottomrule
    \end{tabular}
    \caption{The connected edges for four methods applied on the correlation matrix $\sigb$ and $\sigb^{*}$. The latent variable is denoted as $W$. Abbreviation are TC for Thresholding on correlation, CDG for Conditional dependence graph with thresholding. The threshold is fixed at 0.2 for TC, CDG and FOCI. For TC wth a threshold of 0.2, the dependence graph is a complete graph and so the edges are not listed. The truncation level of the vine is 2 (higher order partial correlations are very small in higher order trees of the vine) in the upper panel and the threshold for partial correlations are set to be 0.2 for drawing edges. The truncation level of the vine is 1 in the lower panel because this suffices to replicate the actual graph of the 1-factor model with each (observed) variable $Z_j$ linked to the latent variable $W$.}
    \label{tab-example-edges}
\end{table}

From Table \ref{tab-example-edges}, for the highly correlated variables based on
a factor model, the graph obtained from the direct-threshold is dense, while the
conditional independency graph becomes sparser as more variables are linked
to the latent variable. 
In this setting, the FOCI and vine methods which relies on the lower-order partial correlations provide more parsimonious dependency graphs.
With the latent variable, the vine dependency graph has an edge for each observed
variables $Z_j$ linked to the latent variable $W$ and this replicates the dependency graph used
in latent variable models and structured equation models; the model \eqref{eq-1factor-gauss} implies $\Cov(Z_j,Z_j|W)=0$ for $i\ne j$ so edges for
conditional dependence are not needed.
Without the latent variable for the graph, the variable most correlated with
$W$ is $Z_1$ and the vine dependency graph first links each of the other
$Z_j$ to $Z_1$; but there is conditional dependence of the other
variables given $Z_1$ --- from 
\eqref{partial-correlations} and 
$\sigb$,
$\rho_{2,3;1}=(0.69-0.77\times0.73)/\rt{(1-0.77^2)(1-0.73^2)}=0.29$ and
$\rho_{9,10;1}=(0.27-0.49\times0.45)/\rt{(1-0.49^2)(1-0.45^2)}=0.06$,
so in Table \ref{tab-example-edges} with a threshold of 0.2, 
the edge [2--3] is added for conditional
dependence but not [9--10].

In the simple 1-factor setting, the FOCI method gives the same graph
as the vine method when the latent variable is included. However, the
FOCI method considers only first-order partial correlation, while the
vine graphs consider the higher-order conditional relationships of the
variables. 

For structured factor models with more than one latent variable (and
structured zeros in the loading matrix), such as data simulated from a
bi-factor model (see Section 3.11.1 of \cite{Joe2014}) with variables
in non-overlapping groups, the FOCI method can not recover the
dependence structure of the latent variable model, and the graph will
be dense to explain the dependence of variables in each group.

One can construct numerical examples with a large $d$ and a structured
bi-factor loading structure with a large number of variables in each
group.  The correlation between observed variables and the global
latent variable can be sampled from a positive interval such as
$[a_{0},b_{0}]=[0.3,0.8]$ while the partial correlations of observed
variables in one group and the local latent variable given the global
latent variable can be sampled from another interval, for example,
$[a_1,b_1]=[0.4,0.7]$, such that the variables in the same group are
more correlated.  The theory in \cite{fan2022high} shows, under some
mild assumptions, that the latent variables can be consistently
estimated from the observed variables in the bi-factor model when the
dimension of each group is large.  Therefore, the partial correlations
of two variables given the remaining variables will decrease towards 0
as $d$ becomes large.  The conclusions for comparison of dependency
graphs of the four methods in the bi-factor case will be similar to
the 1-factor case.

\subsection{Background on truncated partial correlation vines}
\label{sec-vine-concept}
The example in the preceding example was use to provide an idea of how the dependence graph
based on vines includes edges based on some larger absolute partial correlations. The mathematics
behind truncated partial correlation vines is summarized in this section.

A \textbf{vine} as a graphic object on $d$ variables consisting of a sequence of linked trees. The first tree summarizes edges of $d-1$ pairs of variables, with the $d$ variables as nodes. Subsequent trees summarize conditional dependencies. The mathematical definition of vine as a sequence of trees is developed in \cite{bedford2002vines}.
$\v$ is a regular vine on $d$ variables, indexed as $1,\ldots,d$, with $\e(\v)=\bigcup_{i=1}^{d-1}\e(T_{i})$ denoting the set of edges of $\v$, if 
\begin{enumerate}
    \item $\v=(T_1,\ldots,T_{d-1})$ [consists of $d-1$ trees]
    \item $T_1$ is a connected tree with nodes $N(T_1)=\{1,2,\ldots,d\}$, and edges $\e(T_1)$. For $l>1$, $T_{l}$ is a tree with nodes $N(T_{l})=\e(T_{l-1})$ [edges in a tree becomes nodes in the next tree];
    \item (Proximity conditions) for $l=2,\ldots,d-1$, for $\{n_1,n_2\}\in \e(T_{l})$, $\#(n_1\Delta n_{2})=2$, where $\Delta$ denotes symmetric difference and $\#$ denotes carnality [nodes joined in an edge differ by two elements]
\end{enumerate}

\noindent \textbf{Example}: 
The terminology in the above definition is illustrated with $d=4$ variables denoted via indices $1,2,3,4$.
Suppose $[1,2], [1,3], [2,4]$ are edges in tree $T_1$.  For tree 2, we
could have two edges $[2,3;1]$ with nodes $[1,2]$ and $[1,3]$, and
$[1,4;2]$ with nodes $[1,2]$ and $[2,4]$.  If $n_1=[1,2]$ and
$n_2=[2,4]$, then $n_1\Delta n_2=\{1,4\}$ with cardinality 2, and
$n_1\cap n_2=\{2\}$; the edge in tree 2 for nodes $n_1,n_2$ is denoted
as $[1,4;2]$ --- note that no edge can be constructed from $[1,3]$ and
$[2,4]$ because if $n_1=[1,3]$ and $n_2=[2,4]$, then
$n_1\Delta n_2=\emptyset$, the cardinality is not 2 and the proximity
condition does not hold.  For tree 3, nodes $[2,3;1]$ and $[1,4;2]$
have edge $[3,4;1,2]$ because the symmetric difference of the two
nodes is $\{3,4\}$ with cardinality 2.

If the edges of the trees in the vine have values $\rho_{a,b}$ for edges $(a, b)$ in tree 1, and $\rho_{a,b;S}$ for edges $e =(a,b;S)$ after tree 1. A $d\times d$ correlation matrix can be parametrized as a partial
correlation vine with $d(d-1)/2$ parameters that are algebraically independent in $(-1,1)$ using
the values of the partial correlations on these edges; see \cite{kurowicka2003parameterization}. There are many different ways to re-parameterize a correlation matrix into a partial correlation vine. For any regular vine structure $\v$, there is a one-to-one relationship between the entries of $d(d-1)/2$ correlations of a positive definite correlation matrices and the set of algebraically independent $(d-1)$ correlations and $(d-1)(d-2)/2$ partial correlations in the partial correlation vine.

The vine in the above example leads to the partial correlation vine with parameters $(\rho_{12}, \rho_{13}, \rho_{24};\\
\rho_{23;1}, \rho_{14;2}, \rho_{34;12})$. Another partial correlation vine with 4 variables has parameters $(\rho_{12}, \rho_{13},\rho_{14};\\
\rho_{23;1}, \rho_{24;1},\rho_{34;12})$.

\textbf{Truncated vines} consist of a parsimonious way for representing the dependence of $d$ variables.  This is usually done constructing the trees by putting pairs of variables with strongest correlations in tree 1; and pairs of variables with strongest partial correlations $\rho_{a;b;S}$ in low-order trees.
An $m$-truncated vines (with $m\ll d$) assumes that the most of the dependencies
among variables are captured by the first $m$ trees $V_{m}=(T_{1},\ldots,T_{m})$ of the vine. The remaining trees have zero or weak conditional dependence (zero remaining partial correlations for an exact $m$-truncated partial correlation vine, and weak partial correlations below a threshold for an approximate $m$-truncated partial correlation vine).\\

\noindent\textbf{Algorithm for truncated partial correlation vine and vine dependency graph}\\
\indent A sequential maximum spanning tree (MST) algorithm can be used to find
truncated partial correlation vines such the partial correlation in
high-order trees are small.  From \cite{kurowicka2006uncertainty}, the
log-determinant of the correlation matrix
$\log \det (\bfr)=\sum_{e\in \e(\v)}\log (1-\rho_{e}^2)$ for any
regular vine structure $\v$, where $\{\rho_{e}\}$ is the set of
algebraically independent correlations and partial correlations on the
edges of the vine.  Good $m$-truncated partial correlation vine should
lead to a correlation matrix that approximates $\bfr$; this means
\begin{equation}
  \label{eq-objective}
  -\sum_{e\in T_1,\ldots,T_m}\log(1-\rho_{e}^2) \approx -\log\det(\bfr),
\end{equation}
as this would imply the the partial correlations in trees
$m+1,\ldots,d-1$ are closed to 0.

For \eqref{eq-objective}, the choice of edge weight is
$-\log(1-\rho_{e}^2)$ for edge $e$ for the sequential MST algorithm
summarized in Section 6.17 of \cite{Joe2014}.  The $(l+1)$-th tree
$T_{l+1}$ is constructed based on the locally optimal tree $T_{l}$.
In general, the algorithm with local optimality at each tree level is
not globally optimal; enumeration over vines is only possible for
$d\le 7$.  To decide on the truncation level $m$, a
\textit{comparative fit index} is proposed in
\cite{brechmann2015truncation}; this is fine for a sample correlation
matrix when the sample size is large enough relative to $d$.
Otherwise one can stop the sequential MST when the absolute partial
correlation in trees start to be below a specified threshold.

{If the optimum over the left-hand side of \eqref{eq-objective} can be
  found, it might have mostly the same edges at that from the MST
  algorithm, but not necessarily in the same tree; this does not
  affect identification of nodes or genes that have strong links to
  many other nodes/genes.}

For a vine dependency graph, two variables are connected if, in a tree in
the truncated vine, they form an edge with absolute partial correlation
that exceeds a threshold. Variables that are in a path with one or a few
variables in between generally have weak conditional independence given
the intermediate variables but have non-neglible dependence.

\section{Methodology}
\label{sec-method}
In data sets with a large number of variables, one can expect that the
variables can be partitioned into non-overlapping or overlapping
groups, and group information can provide insight on the dependence of
the variables.  We describe the methodology of finding groups and
adding group-based latent variables to get a dependency graph based on
truncated vines.
The outline of the steps are first presented, followed by details of the implementation.

If the variables are not normally distributed, then the first processing step is a rank transform
to standard normal $N(0, 1)$ via the probability integral transform.

Suppose there are $d$ variables (such as gene expression measurements for d genes) and the
sample size is $n$. The variables are assumed to be monotonically related so that summarization
via correlations is meaningful. Let the data vectors be $(x_{i1},\ldots,x_{id})$, for $i=1,2,\ldots,n$. For a fixed variable index $j$,
\begin{equation*}
    z_{ij}=\Phi^{-1}\bigg(\frac{\text{rank}(x_{ij})-0.5}{n}\bigg)
\end{equation*}
where $\Phi$ is the cumulative distribution function of the standard normal distribution. The transformed variables are denoted as $(z_{i1},\ldots,z_{id})$, for $i=1,2,\ldots,n$ and they are said to be on the z-scale. After the transformation, the correlation matrix $\bfr_{\text{data}}$ is obtained from the$(z_{i1},\ldots,z_{id})$ vectors. 

\subsection{Outline of the main steps}
\label{sec-outline}

This section outlines the steps to find groups and introduce latent
variables to explain dependence within groups. Since all genes
interact with other genes in order to form a larger protein complex,
as machinery consisting of small parts, the activities of genes within
the same complex are co-expressed and become present in the cell
simultaneously. Functionally homogeneous groups of genes (variables)
naturally emerge in biological networks, and many empirical studies
indeed confirm the existence of group structures intact in both
manifested gene expression profiles and underlying physical
interaction networks (e.g., \cite{barabasi2004network};
\cite{jansen2002relating}).

To identify groups, some variable clustering algorithms can be applied
to obtain initial non-overlapping groups, and some additional
diagnostic tools to make adjustments so that dependence within each
smaller group has the structure of 1-factor with residual dependence.
For each such resulting group, a latent variable can explain most of
the dependence among variables in the group, and a proxy variable is
created as an estimate of the latent variable.

The procedure consists of several steps.
\begin{enumerate}
\item Split the variables into non-overlapping homogeneous groups
  using a variable clustering algorithm. Suppose there are $m$ groups,
  denoted as $G'_1,\ldots,G'_m$.
\item Because each variable is assigned to some group, a check is made to separate out variables
  that are not strongly related within their assigned group (or other
  groups).  Such variables are considered as isolated from any group.
  There are now smaller groups $G_1,\ldots,G_m$ and a set of isolated
  variables.
\item Now each group $G_g$ has variables with moderate to strong
  correlations.  Fit a 1-factor model to the correlation matrix of
  group $G_g$, and note the variables within the group that can have
  stronger residual dependence.  The remaining variables in $G_g$ have
  the structure of 1-factor with weak residual dependence.

\item For the variables in $G_g$ with weak residual dependence, a
  latent variable $W_g$ can explain most of the dependence, and theory
  from \cite{krupskii2022approximate} provides an approach to compute
  a proxy variable ${\widetilde w}_g $ that estimates the latent
  variable.
\item Add ${\widetilde w}_1,\ldots,{\widetilde w}_m$ to the $d$ variables and 
construct a truncated vine dependency graph using the algorithm
summarized earlier. 
\end{enumerate}

For simpler interpretation, it might be useful to re-orient variables
within each group that are negative correlated with most other
variables in the group. These variable might be those with negative
loadings in the fitted 1-factor mdoels.  A heatmap can be plotted to
visualize the group structure before and after negation of some
variables (on the transformed z-scale).

\subsection{Details of the steps}
\label{sec-details}

\begin{enumerate}

\item For high-dimensional data, a variable clustering algorithm such
  as the CLV algorithm in \cite{vigneau2003clustering} (R library
  \texttt{ClustVarLV}) can be used to form non-overlapping homogeneous
  groups.  The CLV algorithm tries to find an optimal partition of the
  variables to maximize the summation of a homogeneity measure of
  variables within resulting clusters.  The measure of homogeneity is
  larger when the variables are more strongly associated with the
  latent component in each cluster.
\item The weakly correlated variables to separate out can be found
  based on the correlation matrices for the groups $G'_1,\ldots,G'_m$
  from the CLV algorithm.
\item For group $G_g$ ($g=1,\ldots,m$) after step 2, let
  $\bfr_{g,\text{data}}$ be the empirical correlation matrix for group
  $G_g$ and let $\bfr_{g,\text{1-factor}}$ be the correlation matrix
  from a fitted 1-factor dependence model. The residual correlation
  matrix is
  $\bm{D}_g=|\bfr_{g,\text{1-factor}}-\bfr_{g,\text{data}}|$.  A
  variable is considered to have stronger residual dependence if in
  the row of $\bm{D}_g$ corresponding to this variable, there is a
  large value (exceeding threshold 1) or the sum of values in this row
  is large (exceeding threshold 2).
\item Suppose there are $d_g$ variables in group $G_g$ with weaker
  residual dependence: $z_{1g},z_{2g},\ldots,z_{d_g,g}$.  The proxy
  variable are defined as
\begin{equation}
  \label{eq-proxy}
  \tildew_{g}=\Phi^{-1}\Bigl(d_g^{-1}\sum_{j=1}^{d_g}z_{jg}\Bigr).
\end{equation}
\cite{krupskii2022approximate} show that this proxy variable can act
as a good estimate of the latent variable under some mild assumptions
of weak residual dependence, when $d_g$ is large enough and the
variables have been oriented to have positive loadings in the 1-factor
model fit.  The variables with stronger residual dependence are still
in group $G_g$ but there is no guarantee that the proxy with these
variables included would be theoretically consistent.
   
\item In the vine dependency graph with proxy variables, one can
  expect that the variables in group $G_g$ to be mostly linked to
  $\tildew_g$ in the first vine tree.  The additional edges beyond the
  first vine tree will explain the local residual dependence after
  being conditioned on the group latent effect.  Overall, the vine
  dependency graph not only shows the dependence or conditional
  dependence between variables but also provides more information on
  the group structure as well as summarizes the latent dependence
  among variables.
\end{enumerate}

\section{Biological Applications}
\label{sec-data-application}

Two gene applications are presented in this section. 
One involves a yeast gene dataset, and another involves a prostate cancer dataset. The method for building the vine dependency graphs with latent variables is applied to some selected genes of the two datasets. The aim is to explore whether the resulting dependency graphs can be informative and match some biological findings in the literature.

\subsection{MAP kinase pathway inference in Yeast Data}
\label{sec-yeast}

The yeast dataset is from  \href{https://www.ncbi.nlm.nih.gov/geo/query/acc.cgi?acc=gse1990}{Gene Expression Omnibus database (accession no. GSE1990)}. 
From the documentation of the dataset, the haploid segregants from a cross between the yeast strains BY4716 and RM11-1a as in \cite{brem2002genetic}. 
This series contains all GSE617 samples, plus 27 additional segregants assayed with the same protocol and the same reference sample as GSE617, consisting of 
$262$ gene expression vectors of approximately 7,000 genes.
Since the actual regulatory network topology is known for mitogen-actived protein kinase (MAPK) signalling pathway in Kyoto Encyclopedia of Genes and Genomes (KEGG) database \cite{kanehisa2010kegg}, we focused on identifying relationships among $d=37$ genes that are known to constitute the MAPK pathway.
The ground truth gene regulatory networks are available in \cite{kelder2012wikipathways}. 

The proportions of missing values for 50\% of the genes are below 3\%, while others have missing proportion between 3\% and 10\%. For missing values in the raw dataset, we use Gaussian bivariate copulas for imputation.  For the variable with missing values, we find a surrogate variable which is highly correlated with the variable.  The non-missing records of both surrogate and the imputed variable are extracted and rank-transformed to $N(0,1)$ distributed variables.  Then a linear regression model is fitted with surrogate variable as predictor and imputed variable as response.  The prediction on the $N(0,1)$ of the imputed variables from the linear model are then converted to the original scale by the inverse rank transform.  The values are imputed at the missing positions.

\begin{figure}[h!]
\includegraphics[scale=0.19]{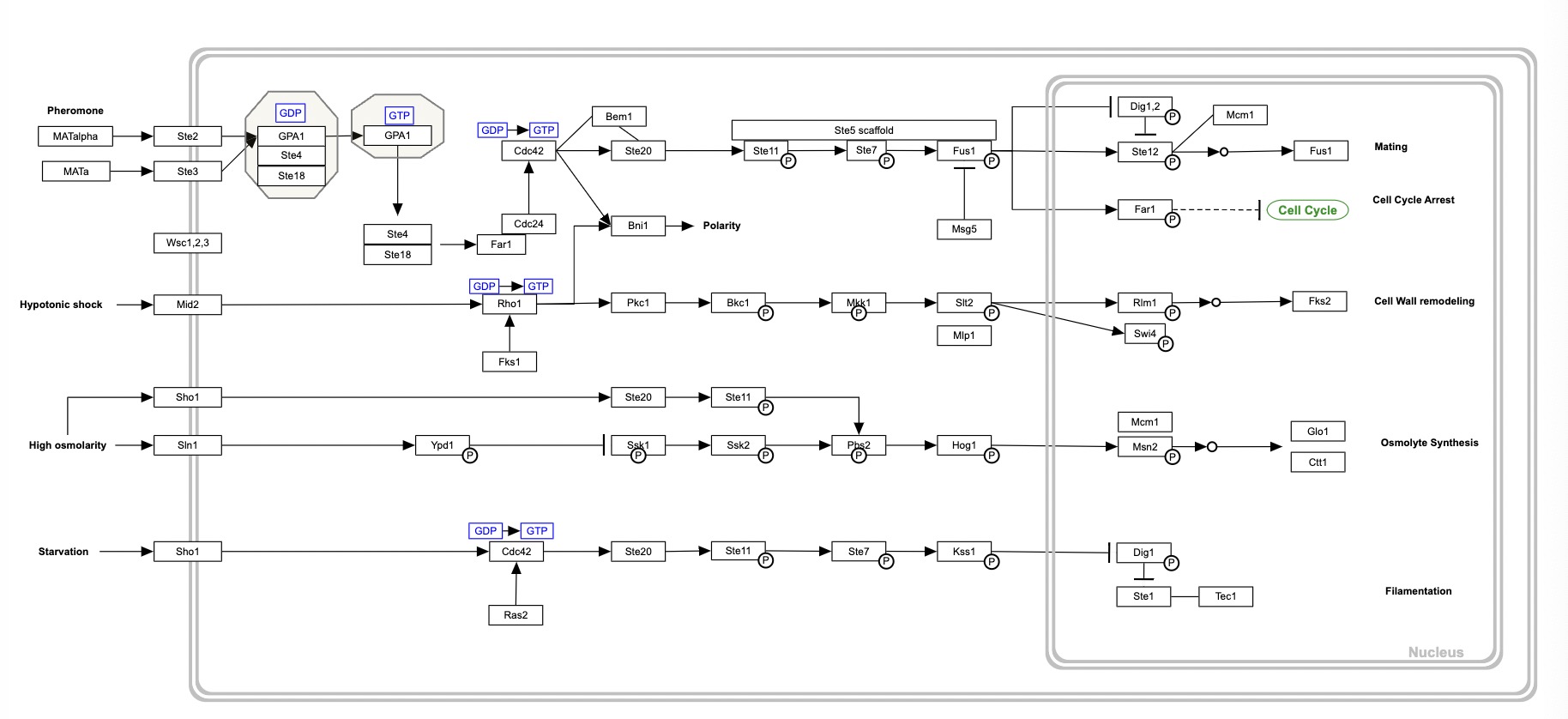}
\caption{MAPK signaling pathway (Saccharomyces cerevisiae) from Wikipedia}
\label{fig-yeast-truth}

\end{figure}

We show the illustrative ground-truth gene regulatory network in Figure \ref{fig-yeast-truth}. In this figure, there are several genes appearing in multiple positions, therefore, it cannot be considered as a dependency graph.
However, our goal is to construct dependency graphs for these genes at the expression level and compare the results with this reference graph to see if there is some biological match.

The proposed methodology 
is applied to the z-scale data after imputation
of missing value and transform to standard normal. 
Combining the initial clustering results with the diagnostic results from the heatmap, there are 4 small homogeneous groups. 
For variables that are mostly negatively correlated with other variables within the groups, we re-orient them by changing the sign such that the correlations within each group are positive. 

\begin{figure}[h!]
\centering
\includegraphics[scale=0.80]{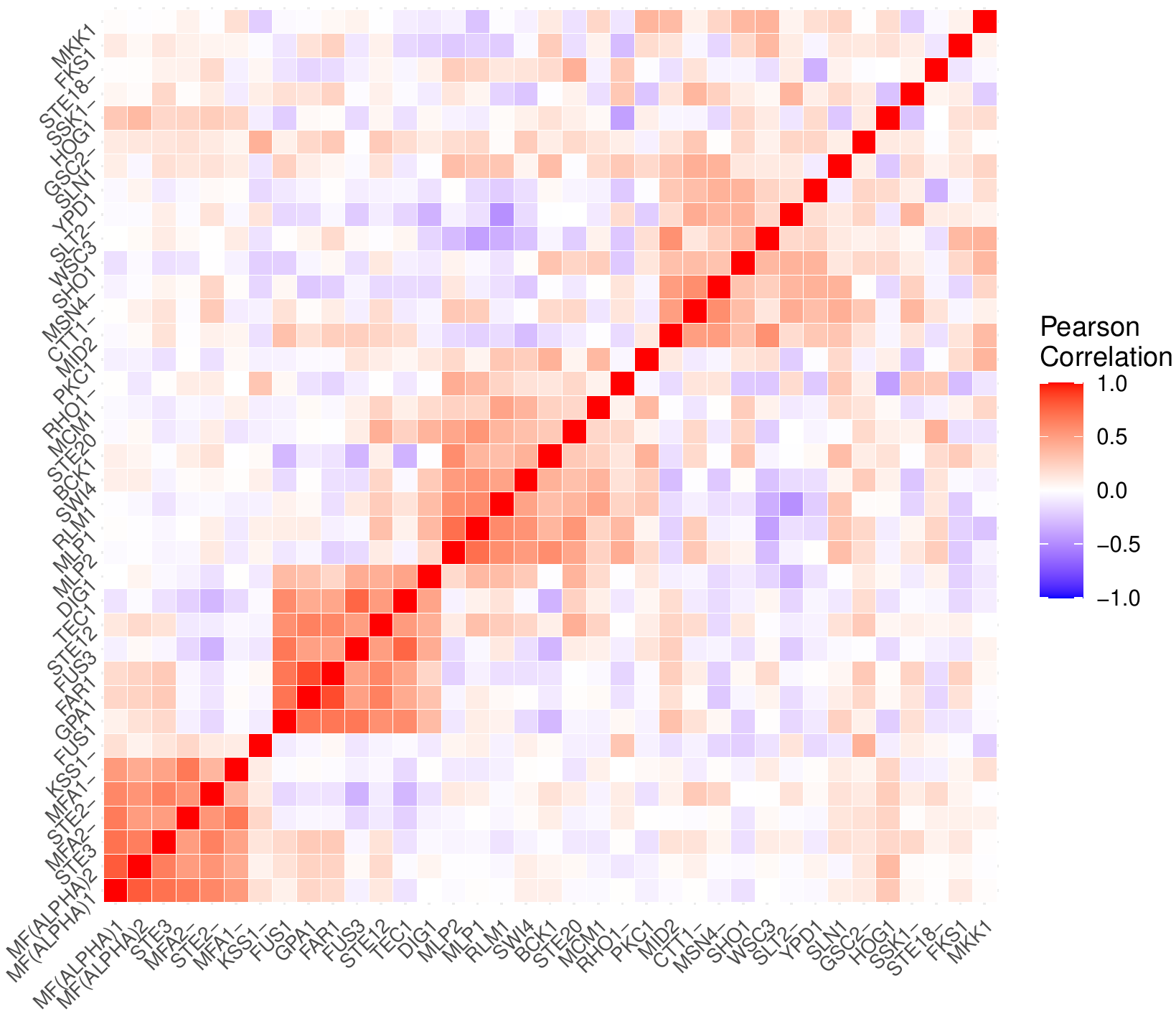}
\caption{Heatmap of the yeast RNA expression data converted to normal scales; there are four homogeous groups from the bottom left to upper right, the group sizes are 7, 7, 9, 8 respectively. There are 6 isolated variables
not in the 4 groups.}
\label{fig-heatmap-yeast}
\end{figure}
\begin{figure}[!ht]
\centering
\includegraphics[scale=0.30]{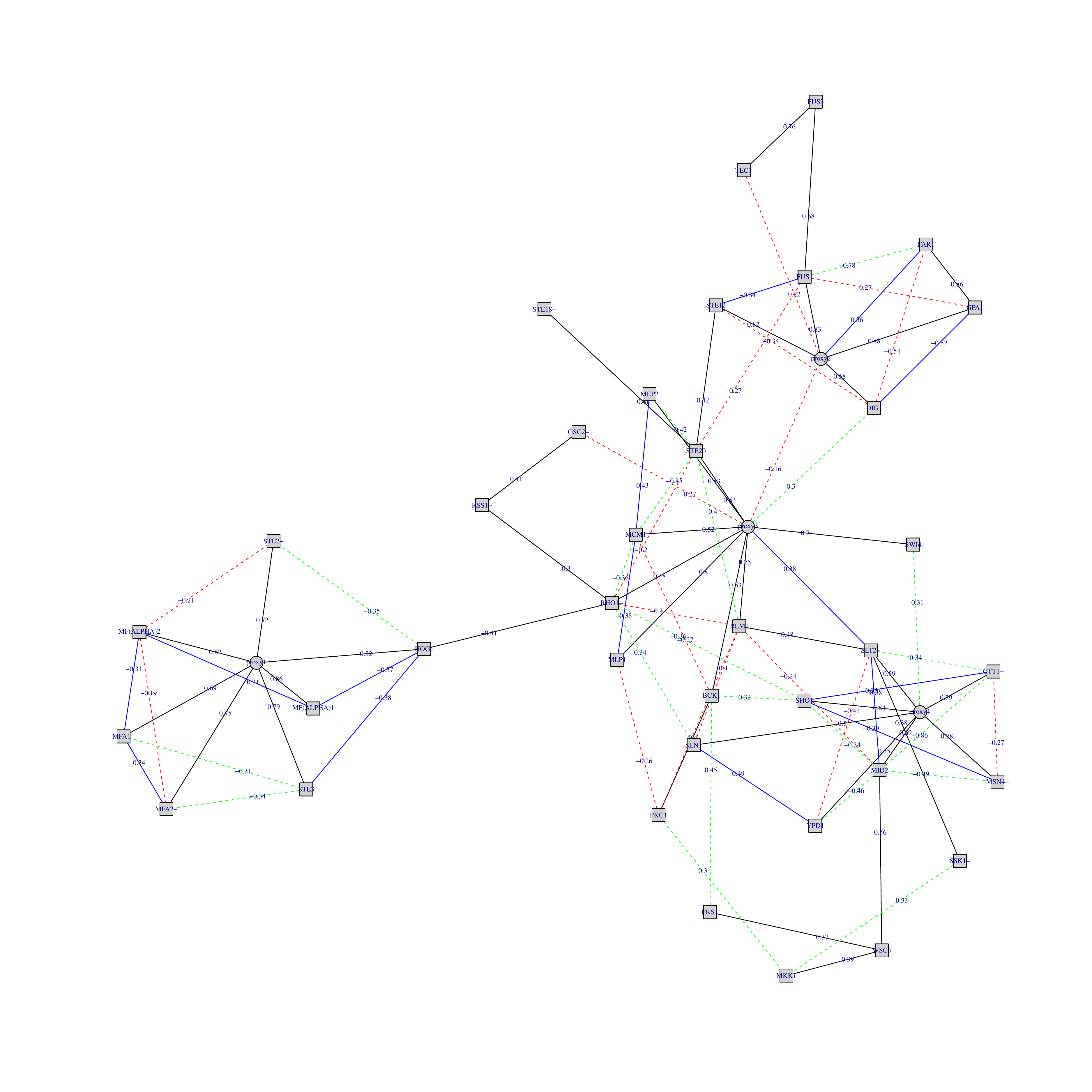}
\caption{\footnotesize{Vine dependency graphs for yeast RNA expression data;  Edges in black from a maximal spanning tree and the connected  edges in blue, red, and green explain the conditional dependence of variables given one, two and 3 or more variables. 
The edges are drawn in the plot if the partial correlation of two variables given one, two and three or more variables are greater than 0.25, 0.15, and 0.30 respectively and are labelled if they are greater than 0.5, 0.3, 0.4 respectively. The labels on the spanning tree in color black is the correlation between two variables while the labels on additional edges are partial correlation of variables given other variables. 
There are four
proxy variables introduced (shown in circles) and the observed variables are shown in rectangles.} }
\label{fig-yeast-dependence}
\end{figure}

For each group, we introduce the proxy variables in \eqref{eq-proxy}
even though the group sizes $d_g$ are not large.
The proxy variables are referred to as proxy1, proxy2, proxy3 and proxy4.
The resulting vine dependency graph is shown in Figure \ref{fig-yeast-dependence}.

Compare the vine dependency graph with the ground-truth figure, the
genes linked to proxy1 and proxy2, such as \textit{MAT-$\alpha$},
\textit{GPA1}, \textit{DIG1}, \textit{FAR1}; they are closely related
and are involved for cell cycle arrest process.  A group of genes,
such as \textit{SLN1}, \textit{SHO1}, \textit{YPD1}, \textit{CTT1},
are linked to the proxy4, and these genes are involved in ``osmolyte
synthesis process''.  Another group of genes, \textit{RHO1},
\textit{BKC1}, \textit{SWI4}, \textit{SLT2}, and \textit{RLM1}, are
linked to the proxy3, and they are involved in ``cell wall modelling
process''.  These findings suggest the links between the proxy
variable and the observed variables could provide some group
information and the genes clustered in the same group are usually
involve in the same cell process.

In addition to the group information, there are some interesting links
in the vine dependency graph matches the ground-truth pathway graph in
Figure \ref{fig-yeast-truth}.  For example, in the cell wall
remodelling process, the genes involved in the process are all linked
closely in the vine graph: \textit{SLT2} linked to \textit{RLM1} in
the first tree and linked to \textit{SWI4} through \textit{RML1}.
This matches with the path \textit{SLT2} $\longrightarrow$
\textit{RLM1} and \textit{SWI4} in the ground-truth pathway.  A group
of genes, \textit{PKC1}, \textit{BCK1}, proxy3 and \textit{RHO1} are
linear in the first tree 1; this roughly matches the ground-truth path
\textit{RHO1} $\longrightarrow$ \textit{PKC1} $\longrightarrow$
\textit{BCK1}.  Furthermore, the genes with many links to the other
genes may act as hub genes and involve in multiple processes.  For
example, \textit{STE20} appears in multiple locations in Figure
\ref{fig-yeast-truth}, and plays roles in multiple cell process. From
the vine dependency graph, it also has many links within two edges to
other genes in different groups.

\subsection{Constructing condition-specific dependency networks in Prostate Cancer Study}
\label{sec-prostate}

We collected gene expression data profiled by RNA-sequencing over $n=497$
prostate adenocarcinoma patients in the cancer genome atlas (TCGA) cohort.
RNA-sequencing technology directly measures the number of short reads mapped
onto each gene (exons), which shows robust correlations with the number of
mRNA molecules transcribed within each gene.  Here, we specifically focus on
genes previously known to be involved in the cancer developmental process
(KEGG), as was done in the previous analysis \citep{lin2016estimation}. For
the selected 314 genes available in the TCGA data, and overlapping with the
KEGG cancer pathway annotation, we applied our proposed methodology.

The CLV clustering algorithm partitioned the total 314 genes into six homogeneous groups, consisting of 55, 44, 63, 62, 44, and 46 variables. In order to estimate 1-factor models, we identified a total of 44 weakly-correlated variables, 3, 6, 5, 15, 8 and 7 from these six clusters and separated them out from each cluster, resulting in the tightly-connected groups of 52, 38, 58, 47, 36, and 39 variables largely explained by a single latent factor within each group.

 For simpler illustration purpose (less busy graphs), we {show the connected sub-graph constructed from the } CLV-groups with index 1, 6, 3 which have more between group dependence.
For groups 1, 6, 3, the numbers of variables used for proxy calculations were respectively 52, 39, 45. The vine dependency graph for all 6 groups in the additional file 1.

From the correlation matrix, some genes have RNA-seq values that
are mostly negatively correlated with RNA-seq values of other genes. 
For these genes, we change the sign of the RNA-seq value in the z-scale;
equivalently this is negation of the original RNA-seq values.
A $-$ (minus sign)
is added to the gene name for the heatmap and vine graphs. 

\begin{figure}[!ht]
\centering
\includegraphics[scale=0.25]{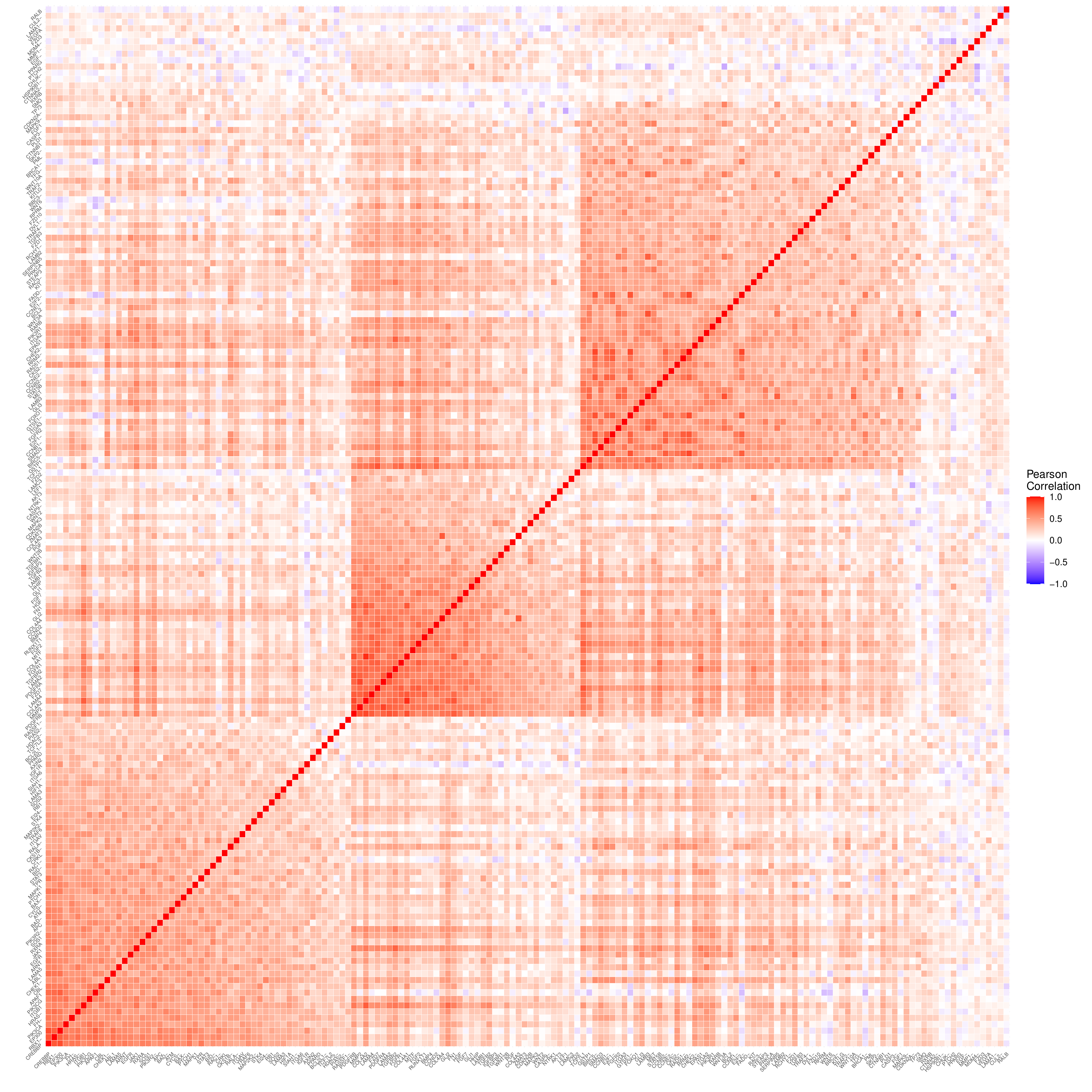}
\caption{A gene expression correlation matrix between 164 genes computed on prostate cancer tumor samples. The color and intensity represent Pearson's correlation values ($-1$ and $1$) based on normalized gene expression z-scores.
From the bottom left to the right, the first block has 52 genes in group 1, the second block has 39 genes in group 6, and the third block has 58 genes in group 3. After third group, there are 15 genes separated out of the CLV groups. }
\label{fig-heatmap-145-tp}
\end{figure}
In Figure \ref{fig-heatmap-145-tp}, there are three homogeneous groups among the 164 genes and we denote the groups along the diagonal from the bottom left to be {group 1, group 6 and group 3}, and then 15 isolated genes
separated from these CLV groups.

For the three groups,
the enrichment analysis is performed on each group. 
The R package \texttt{goseq} is used to identify the ontology of the three groups and the top three potential ontology for each group are shown in Table \ref{tab-ontoplogy}.

\begin{table}[H]
    \centering
\begin{tabular}{c|l}
\toprule
   group & ontopology \\
   \hline
    {1} &   {positive regulation of peptide hormone secretion }\\
    & {drug metabolic process}\\
    & {generation of precursor metabolites and energy}\\
    \midrule
    {6} & {collagen-containing extracellular matrix}\\
    &  
    {angiogenesis}\\
    & %
    {skeletal system development}
    \\
    \midrule
   {3} &  {regulation of muscle system process } \\
   & {nuclear lumen} \\
   &{ nucleoplasm   } \\

    \bottomrule
\end{tabular}
\caption{The top 3 ontologies for the CLV-groups {1, 6, 3} .}
\label{tab-ontoplogy}
\end{table}

\subsubsection*{Vine dependency graphs}

\begin{figure}[!ht]
\centering
\includegraphics[scale=0.33]{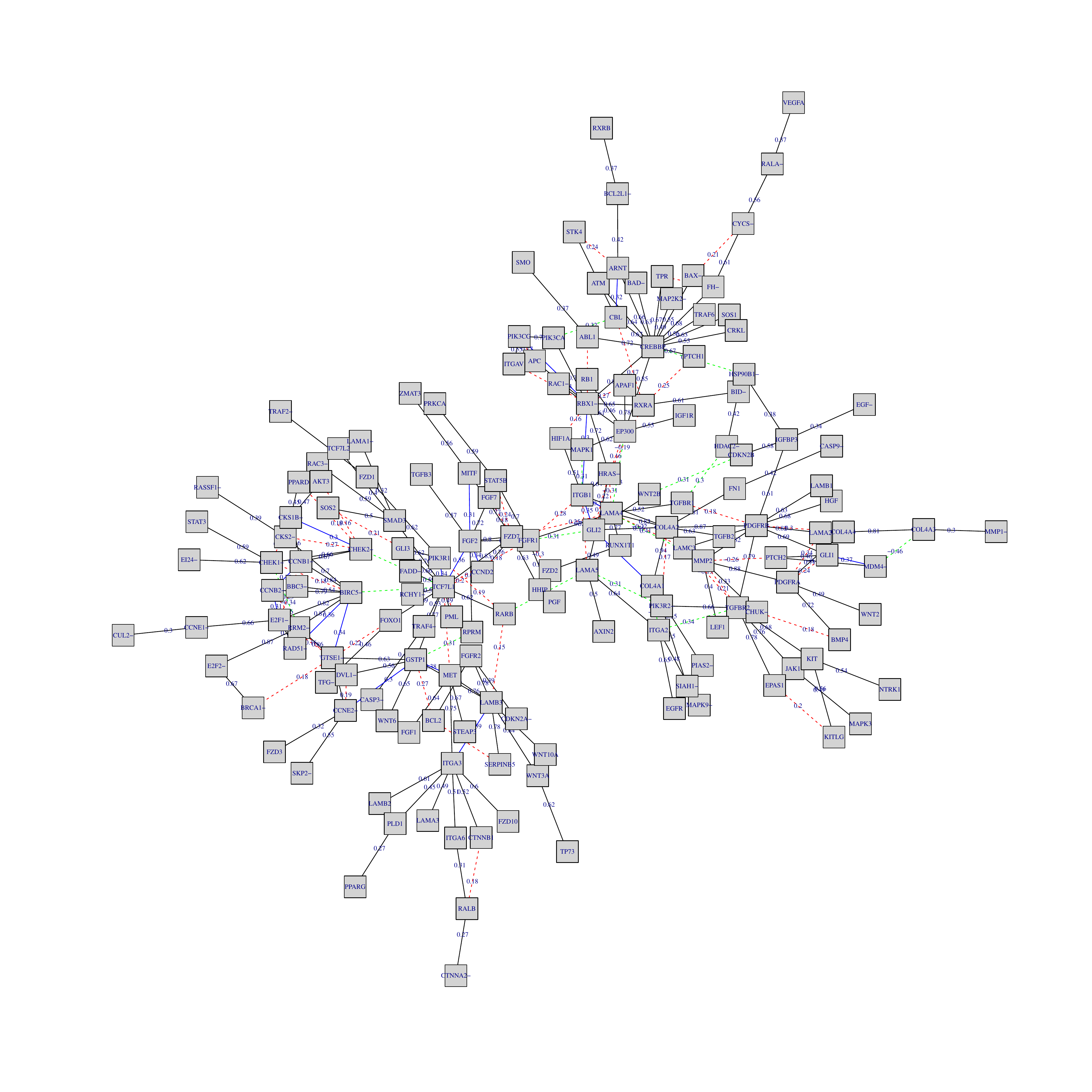}
\caption{\footnotesize{dependency graph for 164 genes in the RNA-seq data of prostate cancer. 
Edges in black from a maximal spanning tree and the connected edges in blue, red, and green explain the conditional dependence of variables given one, two and 3 or more variables. 
The edges are drawn in the plot if the absolute partial correlation of two variables given one, two and three or more variables are greater than 0.30, 0.15, and 0.30 respectively and are labelled if they are greater than 0.5, 0.3, 0.4 respectively. The labels on the spanning tree in color black is the correlation between two variables while the labels on additional edges are partial correlation of variables given other variables. 
There are no proxy variables introduced and the observed variables are shown in rectangles.}}
\label{fig-vine-without-proxy}
\end{figure}

\begin{figure}[!ht]
\centering
\includegraphics[scale=0.33]{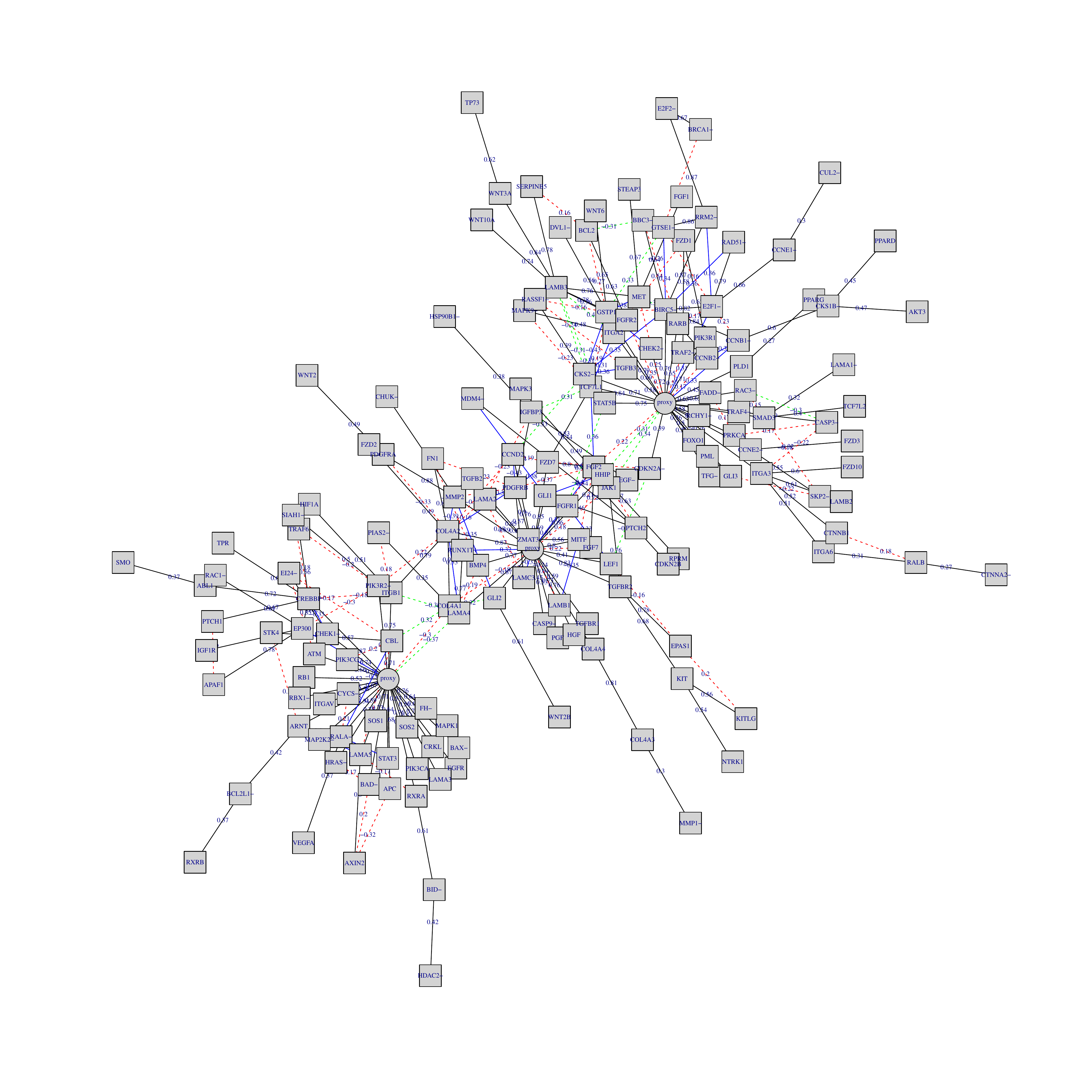}
\caption{\footnotesize{dependency graph for 164 genes in the RNA-seq data of prostate cancer with three proxy variables. 
Edges in black from a maximal spanning tree and the connected edges in blue, red, and green explain the conditional dependence of variables given one, two and 3 or more variables. 
The edges are drawn in the plot if the absolute partial correlation of two variables given one, two and three or more variables are greater than 0.30, 0.15, and 0.30 respectively and are labelled if they are greater than 0.5, 0.3, 0.4 respectively. The labels on the spanning tree in color black is the correlation between two variables while the labels on additional edges are partial correlation of variables given other variables. 
There are no proxy variables introduced and the observed variables are shown in rectangles.}}
\label{fig-vine-with-proxy}
\end{figure}

For the tumor cases, vine dependency graphs were obtained without and with introducing the proxy variables; they are in 
Figures \ref{fig-vine-without-proxy} and \ref{fig-vine-with-proxy} respectively. 

\begin{figure}[!ht]
\centering
\includegraphics[scale=0.5]{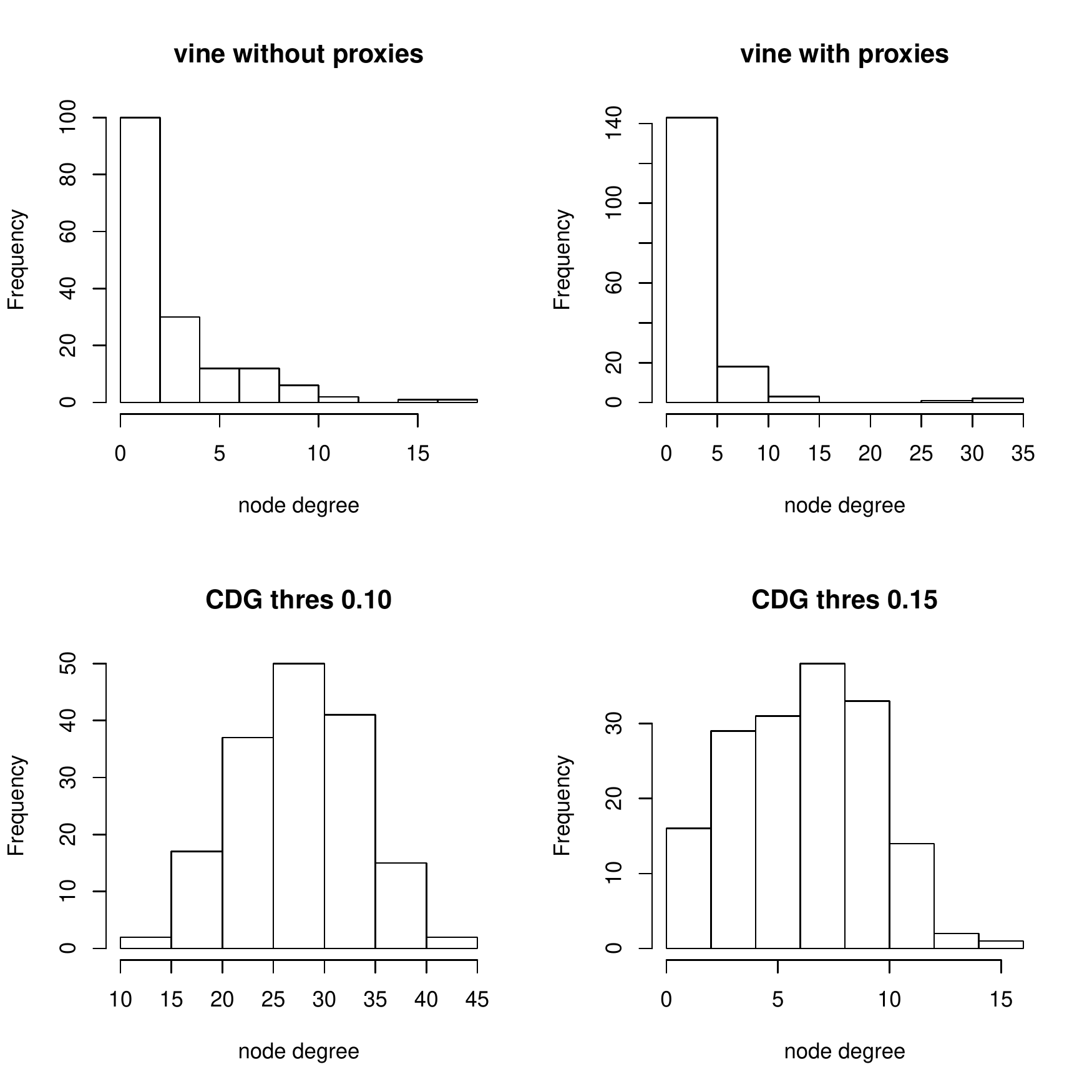}
\caption{Node degree distributions for inferred networks for considered methods.}
\label{fig-node-distribution}
\end{figure}

The simple thresholding on elements of the partial correlations in 
\eqref{eq-partcor-given}
are also performed to get conditional independency graphs. The summary
statistics on the partial correlations are minimum: -0.34, 1st Quartile:
-0.04, Median: 0.00, 3rd Quartle: 0.05, maximum: 0.77. 
Most of the partial correlation are small in absolute value and close to zero. 
 The node degrees distributions for inferred networks for truncated vines
and conditional independency graphs based on different thresholds 0.1 and
{0.15} respectively are shown in Figure \ref{fig-node-distribution}.
{In looking at genes with the largest node degrees and their neighbors,
the resulting subsets are not related to the gene groups based on our
methodology.}
\cite{lin2016estimation} include several methods for producing graphs based
on the inverse covariance matrix; however, we cannot match the tuning
parameters of the methods to thresholds on partial correlations so we cannot
make comparisons. 

Based on the vine dependency graphs, we identify the most connected genes and summarize the results in Table \ref{tab-hub-genes}. 
In the table, we also show the effect size of these genes computed from the
raw RNA-seq counts in log scale and the logFC value (log of fold change)
obtained from the DEG (differential expressed gene analysis) using R package
\texttt{edgeR}.
The fold change describes the differences of RNA expression in gene between normal and tumored cases. The positive fold changes are up-regulated and negative fold changes are down-regulated genes compared to the normal cases.
In addition, all the genes in the table pass the tests for determining
differential expression using the likelihood ratio test in \texttt{edgeR}, when the cut-off p-value is set at 0.05. 

\begin{table}[!ht]
\centering
\caption{\footnotesize{The genes which pass the DEG test (perform LRT test using package ``edgeR''; p-value cut off 0.05) and also have large number of links in the built vine dependency graphs. 
The panel shows the hub genes from the vine graphs built on the tumor samples; the left panel is for vine built without the proxies while the right panel is for vine built with introducing the proxies. effect size is computed in the log-scale: $(\overline{\log y}_{\text{tumor}}-\overline{\log y}_{\text{normal}})/s_{\text{pooled}}$, $y$ is the raw RNA-seq data, logFC denotes the log transform of fold-change which is defined as the ratio of `expression level' in two groups.
Proxy1, proxy6, proxy3 are for groups 1, 6, 3.
For gene names ending with $-$, their RNA-seq values were re-oriented for drawing the heatmap and vine graphs.
}}
\begin{tabular}{lrr|lrr}
\toprule
gene(node-degree)& \multicolumn{1}{r}{effect size}  
& \multicolumn{1}{r|}{logFC} & gene(node-degree)[group] &effect size& logFC     \\ 
\hline
TCF7L1(18) & \multicolumn{1}{r}{-1.81}  & \multicolumn{1}{r|}{-0.86} & proxy3(33)                         &      - &   -    \\ 
CREBBP(16) & \multicolumn{1}{r}{-0.27}  & \multicolumn{1}{r|}{0.23} & proxy1(32)                         &      - &   -    \\ 
RBX1-(11)& \multicolumn{1}{r}{0.43}  & \multicolumn{1}{r|}{0.63} & proxy6(28)                         &      - &   -    \\ 
LAMA4(10)& \multicolumn{1}{r}{-0.86}  & \multicolumn{1}{r|}{-0.45} & COL4A2 (14) [B]                    & -0.78 & -0.51 \\ 
BIRC5-(10)& \multicolumn{1}{r}{1.74}  & \multicolumn{1}{r|}{2.33} &  GSTP1(12)  [C]
&-2.04 & -1.87 \\ 
ITGB1(9)& \multicolumn{1}{r}{-1.28}  & \multicolumn{1}{r|}{-0.51} &  BIRC5-(11)  [C]                    & 1.74 & 2.33 \\ 
FGFR1(9)& \multicolumn{1}{r}{-1.10}  & \multicolumn{1}{r|}{-0.54} &  FGF2(9)   [B]                   & -1.19 & -0.61 \\
FZD7(9)& \multicolumn{1}{r}{-1.21}  & \multicolumn{1}{r|}{-0.64} &  TCF7L1(9)  [C]     & -1.81 & -0.86 \\
GSTP1(9)& \multicolumn{1}{r}{-2.04}  & \multicolumn{1}{r|}{-1.87} & CREBBP(9)   [A]                   & -0.27 & 0.23 \\
COL4A2(8)& \multicolumn{1}{r}{-0.78}  & \multicolumn{1}{r|}{-0.51} &  LAMB3(9) [C]  & -1.28 & -1.42 \\
 FGF2(8)& \multicolumn{1}{r}{-1.19}  & \multicolumn{1}{r|}{-0.61} &  E2F1-(8) [C]   & 0.90 & 1.36 \\
EP300(7)& \multicolumn{1}{r}{-0.25}  & \multicolumn{1}{r|}{0.26} &  MMP2(7)  [B]                    & -0.73 & -0.26\\
LAMB3(8)& \multicolumn{1}{r}{-1.28}  & \multicolumn{1}{r|}{-1.42} & CKS2-(7)  [C]         
& 0.85 & 1.11\\
TGFBR2(8)& \multicolumn{1}{r}{-1.05}  &  \multicolumn{1}{r|}{-0.46} &  CBL(7) [A]
& -0.42 & 0.15\\
ITGA3(8)& \multicolumn{1}{r}{-1.45}  &\multicolumn{1}{r|}{-0.67} &
-  & - &-\\
HRAS-(7)& \multicolumn{1}{r}{0.30}  & \multicolumn{1}{r|}{0.52} & - & -& -\\
GTSE1-(7)& \multicolumn{1}{r}{1.70}  & \multicolumn{1}{r|}{2.02} & -& -\\

RRM2-(7)& \multicolumn{1}{r}{1.79}  & \multicolumn{1}{r|}{1.94} & -  & - & -\\
E2F1-(7)& \multicolumn{1}{r}{0.90}  & \multicolumn{1}{r|}{1.36} & -   & - & -\\
CHEK2-(7)& \multicolumn{1}{r}{1.06}  & \multicolumn{1}{r|}{1.04} & -   & - & -\\

\bottomrule
\end{tabular}
\label{tab-hub-genes}
\end{table}

From the two vine dependency graphs, isolated variables usually have
node degree 1, and these are far from the clusters connected around
the proxy variables.  The genes used to calculate the proxy variables
are usually directly linked to the corresponding group-based proxy
variable, while some of them are linked via a secondary tree to the
proxy variable through one of the other genes in that group.

The hub genes are usually strongly correlated with the proxy variable
in the group and also have a relatively strong correlation with other
variables in the group.  Examples are genes \textit{BIRC5},
\textit{GSTP1}, \textit{TCF7L1}, \textit{CREBBP}; they are linked to a
proxy variable and are connected to other variables in a star shape.
Some hub genes such as \textit{FGF2} connect to proxy variables in one
group and also to genes in another group, acting as a bridge between
the two groups.  Another example for bridging is the gene
\textit{COL4A2}; it connects to the proxy variables of the group {6}
and to the genes of group {1}.  \textit{LAMB3} is a gene from group
{3}, but was not used to calculate the proxy variable because it shows
strong residual dependence.  The graph shows that this gene is linked
to multiple groups.  The links with genes in group {3} explains the
residual local dependence after taking into account the group effect,
while the links with the group {6} proxy variable through the gene
\textit{TCF7L1} explain between-group dependence (and non-weak
residual dependence).

From Table \ref{tab-hub-genes}, there are several hub genes (with
``$-$" after the gene name) that are re-oriented because negative
correlation of RNA-seq values with those of many other genes; examples
are \textit{RBX1}, \textit{BIRC5}, \textit{E2F1}.  The effect size of
these genes are positive, which indicates they are up-regulated in the
prostate cancer patients relative to non-diseased subjects.  The
RNA-seq variables that were not re-oriented had mainly negative effect
sizes, indicating that the corresponding genes are down-regulated in
prostate cancer patients.

Combining the two vine dependency graphs with Table
\ref{tab-hub-genes}, we expect the most connected genes with a
relatively large effect size may play an important role in the
development of the prostate cancer. A shortlist of hub genes in tumor
patients consists of \textit{COL4A2}, \textit{GSTP1}, \textit{BIRC5},
\textit{FGF2}, \textit{TCF7L1}, \textit{CREBBP}, \textit{LAMB3},
\textit{E2F1} from the vine graphs with proxies, and the most
connected genes with large effect size are \textit{RBX1},
\textit{LAMA4}, \textit{ITGB1}, \textit{FGFR1}, \textit{FZD7},
\textit{EP300}, \textit{TGFBR2} in the vine graphs without proxies.

For those potential hub genes, some of them have some related
biological evidence to support them.

\begin{itemize}
\item \textit{TCF7L1}: \cite{wen2021tcf7l1} suggest that induction of
  WNT4/TCF7L1 results in increased NED and malignancy in prostate
  cancer that is linked to dysregulation of androgen receptor
  signaling and activation of the IL-8/CXCR2 pathway.
    
\item \textit{BIRC5}: \textit{BIRC5} is an immune-related gene that
  inhibits apoptosis and promotes cell proliferation.  It is highly
  expressed in most tumors and leads to poor prognosis in cancer
  patients. \cite{xu2021birc5} shows BIRC5 was significantly
  correlated with multiple immune cells infiltrates in a variety of
  tumors.
    
\item \textit{GSTP1}: \cite{martignano2016gstp1} states the gene
  belongs to the GSTs family, a group of enzymes involved in
  detoxification of exogenous substances and it also plays an
  important role in cell cycle regulation. Its dysregulation
  correlates with a large variety of tumors, in particular with
  prostate cancer.
    
\item \textit{FGF2}: From \cite{giri1999alterations} and
  \cite{polnaszek2003fibroblast}, Fibroblast growth factor (FGF) 2 (or
  basic FGF) is expressed at increased levels in human prostate
  cancer. FGF2 can promote cell motility and proliferation, increase
  tumor angiogenesis, and inhibit apoptosis, all of which play an
  important role in tumor progression.  Recently,
  \cite{liu2020identification} also identifies the FGF2 gene as a hub
  gene in the development of the prostate cancer.

\end{itemize}

Based on all the biological evidence, the hub genes we identified from
the small subset (164 genes) play important roles in the progression
of prostate cancer; this indicates that the vine dependency graphs can
provide some interesting and useful insights.

\section{Discussion}
\label{sec-discussion}

This paper presents a method to construct dependency graphs via
truncated vines with latent variables. It makes use of low-order
partial correlations and avoids the problem of high-order partial
correlations used in conditional dependency graphs. Low-order partial
correlations are easier to interpret for the conditional dependence
between variables. With latent variables that can explain much of the
dependence in groups of variables, high-order partial correlations get
closer to 0 in absolute value as the number of varables linked to each
latent variable increases.

Here we address the problem of finding a specialized from of
dependency structures from data while constraining candidate graphs to
be a vine graph. Finding a statistical dependency network from
observed data has long been of great interest to computational
biologists since the first introduction of microarray technology to
date. However, structural learning over combinatorial spaces of
general graphs, such as directed acyclic graphs (DAG), makes exact
latent structure inference highly intractable and resorts to greedy
optimization algorithms \cite{friedman2000using}. Even if we could
enumerate all of these structures, given the limited sample size of
expression data, a scoring function (objective) often run into a
critical problem of distinguishing multiple equally, or at least
similarly, likely structures \cite{chickering2002learning}. However,
interpretable truncated vine structures can be obtained with a
relatively smaller data set.

Statistical dependency structures generally do not correspond to
physical protein-protein interaction networks; hence, care should be
taken for biological interpretations. Instead, dependency maps
typically represent the notion of functional modules, and hub nodes
(high-degree vertices) implicate essential genes that participate in
multiple biological processes. Modeling statistical dependency
structures over high-dimensional gene expression data also benefits
subsequent analysis. For instance, conditional probability calculation
facilitated by a vine graph (conditional independency structures) will
improve regression and classification problems in predicting the
outcome of diseases based on observed gene expressions
[\cite{friedman1997bayesian}, \cite{chuang2007network}]. We could also
expect our approach will provide an efficient way to compute marginal
probabilities over many genes thanks to the sparsity by
construction. This is particularly appealing for G-formula computation
in causality inference \cite{naimi2017introduction}.

\bibliographystyle{abbrv}
\bibliography{bmc_article}

\end{document}